\documentclass[useAMS,usenatbib]{mn2e}

\usepackage{amsmath,array,amsfonts}
\usepackage{amssymb}
\usepackage{graphicx}
\usepackage{placeins}
\usepackage{float}
\usepackage{natbib}
\usepackage{pstricks}
\usepackage{verbatim}
\usepackage{subfigure}

\newcommand{\ltsima} {$\; \buildrel < \over \sim \;$}
\newcommand{\gtsima} {$\; \buildrel > \over \sim \;$}
\newcommand{\lta} {\lower.5ex\hbox{\ltsima}}
\newcommand{\gta} {\lower.5ex\hbox{\gtsima}}

\newcommand{\be}{\begin{equation}}
\newcommand{\ee}{\end{equation}}

\title{Size magnification as a complement to Cosmic Shear}

\author[B. Casaponsa et al.]{B. Casaponsa,$^{1,2}$\thanks{e-mail:
casaponsa@ifca.unican.es}, A. F. Heavens$^3$\thanks{e-mail:
    a.heavens@imperial.ac.uk}, T. D. Kitching$^{4,5}$\thanks{e-mail: t.kitching@ucl.ac.uk}, L. Miller$^6$, R.B. Barreiro$^1$,\newauthor E. Mart\'inez-Gonz\'alez$^1$
 \\
$^1$     Instituto de F\'isica de Cantabria, CSIC-Universidad de Cantabria, Avda. de los Castros s/n, 39005 Santander, Spain.\\
$^2$     Dpto. de F\'isica Moderna, Universidad de Cantabria, Avda. de los Castros s/n, 39005 Santander, Spain.\\
$^3$     Imperial Centre for Inference and Cosmology, Imperial College, Blackett Laboratory, Prince Consort Road, London SW7 2AZ
U.K.\\
$^4$    SUPA, Institute for Astronomy, University of Edinburgh, Blackford Hill, Edinburgh EH9 3HJ, U.K.\\
$^5$    Mullard Space Science Laboratory, University College London, Holmbury St Mary, Dorking, Surrey RH5 6NT, U.K.\\
$^6$    Department of Physics,  University of Oxford, Denys Wilkinson Building, Keble Road, Oxford OX1 3RH, U.K.}
\date{Accepted ;  Received ; in original form }
\begin{document}
\maketitle

\begin{abstract}
We investigate the extent to which cosmic size magnification may be
used to complement cosmic shear in weak gravitational lensing surveys,
with a view to obtaining high-precision estimates of cosmological
parameters.  Using simulated galaxy images, we find that unbiased
estimation of the convergence field is possible using galaxies with
angular sizes larger than the Point-Spread Function (PSF) and
signal-to-noise ratio in excess of 10. The statistical power is similar to,
but not quite as good as, cosmic shear, and it is subject to different
systematic effects.  Application to ground-based data will be
challenging, with relatively large empirical corrections required to account for 
the fact that many galaxies are smaller than the PSF,
but for space-based data with 0.1-0.2 arcsecond resolution, the size
distribution of galaxies brighter than $i\simeq 24$ is almost ideal for
accurate estimation of cosmic size magnification.  
\end{abstract}
\section{Introduction} 

General relativity predicts that the path of light from a distant galaxy is distorted by the gravitational potential fluctuations along the line of sight.
This modification of the light paths is called gravitational lensing and is a powerful tool for probing the distribution of mass in the Universe.
The variation of the light path depends on the position in the sky of the emitting object, the distance from the emitting object to the
observer and on the potential along the light path.
As the Universe is in permanent evolution photons emitted at an earlier epoch will be deflected 
from those emitted later, due principally to the longer path length  \citep[for some reviews see][]{Schneider1992,Narayan1996,Mellier1999,Munshi2008}. 
Combining this depth information with angular information on the gravitational lensing distortion, allows for an unbiased reconstruction
of the three-dimensional distribution of matter. In addition, statistical analysis to infer cosmological parameters can be made from the 
dependence of observables on the distance-redshift relation and growth 
of density perturbations with redshift. Dark energy and modifications to Einstein gravity also act to modify the lensing effect by changing the 
distance-redshift relation in addition to the growth of density perturbations 
\citep{Huterer2002,Munshi2003}. Lensing effects are therefore a particularly valuable source of information for three of the important open
issues in modern cosmology, namely the distribution of dark matter, the properties of dark energy and the nature of gravity.

Observationally there are three main effects on the background sources: changes in the ellipticity, magnification of the flux, 
and magnification of the size, the last two being directly related due to the conservation of the surface brightness in gravitational lensing. For large densities of matter the effects are all very strong, and multiple images of the background 
galaxies can be produced with large distortion. The studies of this distortion have led to local reconstructions of the distribution of 
matter \citep[see for example][]{Tyson1984,Fort1988,Tyson1990,Kaiser1993,Mellier1993}.
When the potential fluctuations 
and their derivatives are small, the mapping from the source position to the image position on the sky is the identity matrix with corrections which are $\ll 1$.  This weak-lensing regime does not allow significant information to be gained from individual sources, but 
the distortions may be observed statistically using a large sample.

In weak lensing, the most studied effect is the modification to the
galaxy shape, a measure of the shear. The shape distortion has the main advantage  
that the intrinsic distribution of galaxy ellipticities is expected to be random, 
according to the cosmological principle, and therefore the average complex ellipticity is zero. 
Weak lensing effects using galaxy ellipticities is a well-developed field and has been detected by several 
groups using different surveys and different methods, see for example \citet{Wittman2000,Sembeloni2006,Jarvis2006,Benjamin2007,Schrabback2010}. 
In addition important efforts have been made to include and test many possible systematic effects on 
shape measurement (including the point spread function [PSF], instrumental noise, pixelization for example), and there are several algorithms that 
can measure shapes with varying degrees of accuracy including KSB \citep{Kaiser1995}, KSB+ \citep{Hoekstra1998} and its 
variants \citep{Rhodes2000,Kaiser2000} and shapelets \citep{Bernstein2002,Refregier2003,Kuijken2006} amongst others. 
A novel Bayesian model fitting approach {\em lens}fit
was presented in \citet{Miller2007, Kitching2010}. In order 
to test, in a blind way, the ability of methods to measure the shapes of galaxies a series of 
simulations have been created: STEP1, STEP2, GREAT08 and GREAT10, where
several methods have been tested and compared systematically. An explanation of the 
methods and their performance are summarised in 
\citet{Heymans2006, Massey2007, Bridle2010, Kitching2012} respectively. 
In addition to altering the shapes of galaxies, weak lensing induces a flux magnification effect, which varies the number of galaxies above a flux threshold, and is usually quantified using the number counts 
(see for example \citealt{Broadhurst1995,Hildebrandt2009,Hildebrandt2011}); 
the number of observed galaxies above a given flux may increase or decrease due to lensing by the 
foreground galaxies increasing fluxes but simultaneously increasing the solid angle covered by the images. We do not study this further in the paper but instead focus on size magnification. 

In contrast to galaxy ellipticity measurement, the size information has not been explored in detail, possibly because the
complicating effects of the PSF and pixellisation were thought to be
too challenging.  However, there are two reasons for revisiting size
magnification as a potential tool for cosmology: one is that accurate
shear estimation is itself very challenging, and 
size could add useful complementary information; the second is
that methods devised for ellipticity estimation must deal with the PSF
and pixellisation, and as a byproduct provide a size estimate, or a
full posterior probability distribution for estimated size, which is currently ignored or marginalised over.  

In terms of signal-to-noise (S/N) of shape distortions vs magnification estimation, the relative strengths of the methods depend on the distributions of ellipticity and size.  The former has an r.m.s. of around 0.3-0.4 \citep{Leauthaud2007}; for bright galaxies ($M_r<-20$), the Sloan Digital Sky Survey (SDSS) found that the size distribution is approximately log-normal with $\sigma \ln R\sim 0.3$, 
and for fainter galaxies $\sigma (\ln R)\sim 0.5$), where $R$ is
the Petrosian half-light radius \citep{Shen2003}; for deeper space data the dispersion is also around 0.3 \citep{Simard2002}. Thus one might
expect a slightly smaller S/N for lensing measurements based on size rather than ellipticity, but not markedly so. 
\citet{Bertin2006} proposed a method based on the fundamental plane relation \citep{Dressler1987,Djorgovski1987} to 
reduce the observable size variance. \citet{Huff2011} applied a similar method 
to measure galaxy magnification using $55$,$000$ galaxies of the SDSS catalogue, 
and find consistency with shear using the same sample.  Also a detection with COSMOS HST survey using a 
revised version of the KSB method is presented in \citet{Schmidt2012}, showing reasonable consistency with shear.   

Here we revisit size magnification measurement, and will show that to use size-magnification we require 
i) a wide area survey that enables observations of a sufficiently large sample of galaxies, this is required to overcome the intrinsic scatter, 
and ii) a consistently small PSF that does not destroy the size information of the observed galaxies. 
Both of these requirements can be met with a wide-area space-based survey, although some science may be possible from the ground. 
Euclid\footnote{{\tt http://www.euclid-ec.org}} \citep{Laurejis2011} should meet these requirements (large samples will be available, and the PSF size is smaller than typical galaxies), so the size information could be considered as a complementary cosmological probe to weak lensing ellipticity measurements. 
One advantage of using the size information is that the magnification and distortion have different radial dependences on the spatial distribution of matter, which may be very useful to lift the so-called mass-sheet degeneracy \citep{Bartelmann1996,Fort1997,Taylor1998,Broadhurst2005,Umetsu2011,Vallinotto2011}, 
that occurs due to the reduced shear (or the measured ellipticities) being invariant under a transformation of the distortion matrix by a scalar multiple.

Besides the degeneracy lifting, 
another advantage of using magnification is that combining the size magnification information with the shear will reduce 
uncertainties on the reconstruction of the distribution of matter \citep{Jain2002,Vallinotto2011,Sonnenfeld2011}.  

On the measurement of the size, all of the shear estimation methods referred to above already estimate the size of galaxies
when calculating the ellipticities, so we expect to measure this additional information for free, given an accurate ellipticity measurement. However, the 
accuracy of size information should not be taken for granted: it is important to know the uncertainties in size measurement, 
and how they propagate to a convergence field estimation. It is this question of how accurately one can measure the sizes of galaxies, that this paper addresses. \citet{Amara2007,Kitching2008a} have shown that to obtain an accurate determination of cosmological parameters, such as the equation of state of dark energy, the systematic errors in the measured ellipticities should be $\lesssim 0.2\%$, and we would expect similar requirements for size. Although a full study of the convergence bias at this level needs to be done, the main goal of this first paper is to investigate whether unbiased measurement of size is feasible at all, and to come some basic conclusions on required image sizes and signal-to-noise. 

The paper is organised as follows. First, in Section 2 we will present 
the weak lensing quantities that are used throughout this paper, then a definition of the estimator, and a brief comment on 
the method and the characteristics of the simulated images. In Section 3  the analysis and results are explained and finally 
we will summarise the conclusions in Section 4.
 
\section{Method}
A good algorithm for weak lensing analysis must be able to 
take into account the distortion introduced by the PSF, pixelization effects and pixel noise. 
Another requirement is that it should be computationally fast because the statistical
analysis will be made on large samples. This means that algorithm development is challenging
because of the dissonant requirements of both increased accuracy and increased speed as the required systematic level decreases. 
Several methods have been proposed and applied to weak lensing surveys, and are described in the challenge reports 
of STEP, GREAT08 and GREAT10 \citep{Heymans2006,Bridle2010,Kitching2012}. These blind challenges have been critical in demonstrating to what extent methods can achieve the required accuracy for upcoming surveys by creating simulations with controlled inputs against which results can be tested. 
Here we propose a very similar approach as the one presented in the GREAT10 challenge, we have used simulated galaxy images with 
different properties to measure the response of the size/convergence measurement under different conditions (corresponding to changes in the 
PSF, S/N and bulge fraction). 
\subsection{Weak lensing formalism}
The distortions induced by gravitational lensing are described by the Jacobian matrix which maps the true angular position of the image to the angular position of the source (in the absence of deflections):
\[ \mathcal A(\vec{\theta})=\left( \begin{array}{c c}
1-\kappa-\gamma_1 & -\gamma_2 \\
-\gamma_2 & 1-\kappa+\gamma_1\\
\end{array} \right)\]
which defines the convergence field $\kappa$ and complex shear field $\gamma\equiv \gamma_1+i\gamma_2$ \citep[for more details, see e.g.][]{Bartelmann1996,Hoekstra2008,Munshi2008}.
In terms of the convergence and the shear there are two important variables, directly related to the lensing observables. The reduced shear, 
\be 
g=\gamma (1-\kappa)^{-1}\label{red_shear}
\ee 
represents shape changes ignoring size.  The magnification of the surface area $\mu$ is, 
\be 
\mu = \frac{1}{\det(\mathcal A)}=[(1-\kappa)^2-|\gamma|^2]^{-1}.\label{mag}
\ee 
If $\kappa,|\gamma|\ll 1$ (which we assume throughout) can be approximated by 
\[
\mu\simeq 1+2\kappa.
\]  
The power spectrum for $\kappa$ can be obtained in terms of the matter power spectrum $P_{\delta}(k,w)$, where $w$ is a comoving distance coordinate which plays the role of cosmic time.  For a set of sources with a distribution function $p(w)$,
\begin{equation}   
  P_\kappa(\ell) = \frac{9H_0^4\Omega_{\rm m}^2}{4c^4}\,
  \int_0^{w_{\rm h}} dw\frac{g^2(w)}{a^2(w)}
  P_\delta\left(\frac{\ell}{w},w\right)\,,
\label{power_spectrum}
\end{equation} 
where $g(w)=\int_w^{w_{\rm h}}dw'\,p_w(w'){w'-w\over w'}$, $w$ is the co-moving distance, $w_{h}$ is the horizon distance. $\Omega_m$ and $H_0$ are the present matter density parameter and Hubble parameter. Note that eq.~\ref{red_shear} and \ref{mag} are always valid while eq.~\ref{power_spectrum} is only meaningful for cosmic shear.

In the weak lensing limit, the power spectrum of the magnification fluctuations ($\mu -1$) is 4 times $P_k(l)$, therefore, in principle, 
cosmological constraints could be made independently of the shear \citep{Jain2002,Barber2003}, however the signal-to-noise ratio for the measured ellipticities 
is in general larger, hence the shear may carry more statistical weight. Even so a complementary analysis of shear and magnification measurements will necessarily provide tighter constraints 
on cosmological parameters than a shear analysis alone. In particular, in \citet{VanWaerbeke2010} it is shown that the constraints on $\sigma_8$ and $\Omega_m$ can be improved up to 
$\sim 40\%$, similarly combining size-magnification, galaxy densities and shear, the improvement on the precision of halo mass estimates can be $\sim 40\%-50\%$ \citep{Rozo2010}. 

\subsection{Estimator} 
In this paper we will work with the semi-major axis of the images, denoted by $s$, and the semi-major axis of the unlensed source,  $s^{s}$.  Their ratio is given by $\frac{s}{s^s}=(1-\kappa-\gamma_1')^{-1}$, where $\gamma_1'$ is computed in the principal axis frame.  At linear order in the weak lensing regime their ratio is $\frac{s}{s^s}=1+\kappa+\gamma_1'$.  The contribution of shear is a zero-mean noise term which is small in comparison to the effects of the intrinsic size distribution, and will be ignored. On the other hand,
the magnification is related to the convergence, to first order, $\mu^{\frac{1}{2}}=1+\kappa$, therefore we can write $\frac{s}{s^s}=\mu^{\frac{1}{2}}+noise$ then $\kappa\sim\frac{s}{s^s}-1$.
 We can construct an
estimator for $\kappa$ in the weak lensing limit by assuming that, since $\langle \kappa\rangle =0$, the 
mean size value is not modified by lensing, i.e., $\langle s^s\rangle=\langle s\rangle$, and replacing $s^s$ by its expectation value: 
\begin{equation}
\hat{\kappa}=\frac{s}{\langle s\rangle}-1. 
\label{estimator}
\end{equation}
From the definition of the estimator, and the width of the $s$ distribution, an estimate of $\kappa$ from a single galaxy will be very noisy, with smaller galaxies than the mean always 
giving a negative $\hat{\kappa}$, while larger galaxies will produce positive $\hat{\kappa}$. 
What is important is to test if our estimator is unbiased over a
population to a sufficient degree to be useful for real data. 

\subsection{{\em lens}fit}
\label{LENSFIT}
Throughout we use {\em lens}fit \citep[][Miller et al., 2012]{Miller2007,Kitching2008} to estimate the galaxy
size; we use this because: 1) it has been shown that {\em lens}fit performs well on ellipticity
measurement; 2) it is a model fitting code which also
measures the sizes of galaxies; 3) it allows for the
consistent investigation of the intrinsic distribution of galaxy sizes through the inclusion of a prior on size, and 4) it includes the effects of PSF and pixellisation.
{\em lens}fit was proposed in \citet{Miller2007} and has been proved to be a successful tool for galaxy ellipticity shape measurements 
\citep{Kitching2008}. Although model-fitting is the optimal approach for this type of problem if the model used is an accurate representation of the 
data, the main disadvantage is that is usually computationally demanding to explore a large parameter space.
{\em lens}fit solves this problem by analytically marginalizing over some parameters that are not of interest for weak lensing ellipticity 
measurements, such as position, surface brightness and bulge fraction. The size reported by {\em lens}fit is also marginalised over the galaxy ellipticity.

\subsubsection{Sensitivity correction}
\label{sens}
In the Bayesian formalism the expected value for the size of an individual galaxy can be written as $\langle s \rangle=\int{s\,\emph p(s|s_d)ds}$, where $s_d$ is the data and $s$ stands for the fitted model parameter for the size explained in Sec.~\ref{simulations}. In terms of the prior $\mathcal P(s)$ and the likelihood $\mathcal L(s_d|s)$ the expression is
 \begin{equation}\langle s\rangle=\frac{\int{s\mathcal P(s)\mathcal L(s_d|s)ds}}{\int{\mathcal P(s)\mathcal L(s_d|s)ds}} .
\label{eq_expvalue}
\end{equation}
Individual galaxy size estimates allow errors to be assigned to each galaxy, or the full posterior can be used and the information propagated to the $\kappa$ signal. \citet{Miller2007} introduced the shear sensitivity, a factor that corrects for the fact that the code measures ellipticities but that shear (a 
statistical change in ellipticity) is the quantity of interest. 
A similar correction is required for size measurement, whereby we measure the size but it is the convergence that is the quantity of interest; 
this correction is needed because for a single galaxy the prior information for the convergence is not known, and we assume it is zero. 
With a Bayesian method we can estimate the magnitude of this effect for each galaxy, a further reason to use a Bayesian model fitting code in 
these investigations. Consider the Bayesian estimate of the size of galaxy $i$ and write its dependence on $\kappa$ as a Taylor expansion:
\begin{equation}
\hat{s}_{i}=s_{i}^s+\kappa\frac{d \hat{s}_i}{d\kappa}.\label{kappasens}
\end{equation}
In the simple case where the likelihood $\mathcal L(s_d|s)$ (for simplicity, hereafter $\mathcal L(s)=\mathcal L(s_d|s)$) is described by a Gaussian distribution with variance $b^2$, with an expectation
value $s$, and a prior $\mathcal P(s)$ that also follows a Gaussian distribution centred on $\bar{s}$ with variance $a^2$, 
the posterior probability will follow a Gaussian distribution with expectation value:
\[
\langle s\rangle=\frac{\bar{s} b^2 + s_d a^2}{a^2+b^2}
\] and variance  \[ \sigma^2=\frac{a^2 b^2}{a^2+b^2}.\]
 
These equations illustrate that the posterior is driven towards the prior in the low S/N limit ($b\rightarrow \infty$), and thus requires correction.  Differentiating the expression $s_d=s^s(1+\kappa)+\sigma_s$, with $\sigma_s$ being the systematic noise, we find that the $\kappa$ sensitivity correction is:
\begin{equation}
\frac{d \hat{s}}{d \kappa}= \frac{a^2}{a^2+b^2}\frac{d s_d}{d \kappa}=\frac{a^2}{a^2+b^2}s^s, \label{kappasens2}
\end{equation}    
substituting into eq.~\ref{kappasens} 
\[ \hat{s}_{i}=s_i^s+\kappa s_i^{s} \frac{a^2}{a^2+b^2}\]
we find the estimator for $\kappa$ will be the same as in eq.~\ref{estimator}, corrected by the sensitivity factor:
\begin{equation}
\kappa=\left(\frac{\hat{s}}{\langle\hat{s}\rangle}-1\right)\frac{a^2+b^2}{a^2}. \label{estimator_corrected}
\end{equation}
In this work we have used this approximation for simplicity but in general a normal distribution should not be assumed. 
A more general estimation of the $\kappa$ correction can be done in the same way as with the shear and can be evaluated numerically, 
without the need of using external simulations. 

To calculate the sensitivity correction in the general case we consider the response of the posterior to a small $\kappa$, by adding the convergence contribution in the likelihood, $\mathcal L(s-s^s) \mapsto \mathcal L(s-s^s-\kappa s^s)$ and expand it as a Taylor series:
\[\mathcal L(s-s^s-\kappa s^s)\simeq \mathcal L(s-s^s)-s^s\kappa\frac{d\mathcal L}{d s}. \]  
We then substitute into eq.~\ref{eq_expvalue} and differentiate to obtain the analytic expression 
for the $\kappa$ sensitivity (for more details of this applied to
ellipticity measurement see \citealt{Miller2007,Kitching2008})
\begin{equation}
\frac{d s}{d \kappa}\simeq \frac{\int{(\langle s\rangle-s)\mathcal P(s)s^s \frac{d\mathcal  L}{d s}}ds}{\int{\mathcal P(s)\mathcal L(s) ds}}.
\end{equation}
If the prior and likelihood are described by a normal distribution, this expression can be analytically 
computed and the sensitivity correction is the same as before. A similar empirically motivated 
correction on the estimator expression was used in eq.5 of \citet{Schmidt2012}, where the factor is computed with simulations.

\subsection{Simulations}
\label{simulations}
In order to test the estimation of sizes with {\em lens}fit we have generated the same type of simulations used in the GREAT10 challenge 
\citep{Kitching2010,Kitching2012}, but with non-zero $\kappa$. Multiple images were generated, each containing 10,000 simulated galaxies 
in a grid of 100x100 postage stamps of $48$x$48$ pixels; each postage stamp contains one galaxy.

Each galaxy is composed of a bulge and a disk, each modeled with S\'{e}rsic light profiles: 
\begin{equation}
I(r)\simeq I_{0}\exp{\left\lbrace-K \left[\left(\frac{r}{r_d}\right)^{\frac{1}{n}}-1\right]\right\rbrace}
\end{equation}
where $I_{0}$ is the intensity at the effective radius $r_d$ that encloses half of the total light and $K=2n-0.331$. 
The disks were modelled as galaxies with an exponential light profile ($n=1$), and the bulges with a de Vaucouleurs profile ($n=4$). 
Ellipticities for bulge and disk were drawn from a Gaussian distribution centred on zero with dispersion $\sigma=0.3$. 
Both components had distributions centred at the middle of the postage stamp with a Gaussian distribution of $\sigma=0.5$ pixels. 
The galaxy image was then created adding both components. The S/N was fixed for all galaxies of the image and implemented by calculating the noise-free model flux by integrating over the galaxy
model, then adding a constant Gaussian noise with a variance of unity and rescaling the galaxy model to yield the correct signal-to-noise, as in \citet{Kitching2012}. 
Finally the PSF was modelled with a Moffat profile with $\beta=3$, with FWHM  
fixed for all galaxies on the image, with different ellipticities drawn from a uniform distribution, with ranges given in Table~\ref{table1}.

The different types of image were generated to study the effects of the bulge fraction (fraction of the total flux concentrated in the bulge), 
the S/N and the PSF separately. In summary, the main characteristics of the considered sets are:
\begin{itemize}
\item Set 1. Disk-only galaxies (bulge fraction = 0), negligible PSF
  effect (FWHM PSF = 0.01 pixels) and different S/N.
\item Set 2. Disk-only galaxies (bulge fraction = 0), with S/N=20 and different sizes of PSF.
\item Set 3. Negligible PSF effect, S/N=20 and different bulge fractions.
\item Set 4. Bulge fraction of 0.5, FWHM of PSF 1.5 times smaller than the characteristic size of the disk, and different S/N.
\end{itemize}
To characterize the size of the galaxy the semi-major axis is used, related to the half-light disk radius $s=\frac{r_{d}}{\sqrt{q}}$, where $q$ is the semi-axis ratio. We have drawn $r_d$ from a Gaussian distribution 
with expected value of $7$ pixels and dispersion of
$1.2$ pixels, to keep disk sizes of at least $2$ pixels and not larger than 
the postage-stamp. The galaxy sizes explored here have a somewhat
smaller range ($\sigma(\ln R)\sim 0.18$) than found by \citet{Shen2003} with the SDSS catalogue, 
where in terms of pixels the mean value of the full sample is around
$5$ with $\sigma(\ln R)\sim 0.3$ 
\citep[see Fig.1 of][]{Shen2003}. Therefore the sensitivity corrections are consequently larger than would be needed for real data.
Besides a wider distribution of galaxies, the important change from the original images for the GREAT10 challenge is the addition of a non-zero $\kappa$-field that
creates a size-magnification effect (in GREAT10 only a shear field was used to distort the intrinsic galaxy images). A Gaussian convergence field 
with a simple power-law power spectrum in Fourier space $P_\kappa(\ell)\sim10^{-5}\ell^{-1.1}$ has been applied to each image. The power-law is a good approximation to the 
theoretical power spectrum over the scales $10\lesssim \ell \lesssim 10000 $ \citep[see i.e.][]{Schneider2005}.The size of the $\kappa$-field is $\theta_{image}= \frac{2\pi}{\ell_{min}}$ and $\theta_{image}$ is set to 10 degrees, such that the range in $\ell$ we used to generate the power was $\ell = [36, 3600]$ where the upper bound is given by the grid separation cut-off. In real space this translates to a maximum $|\kappa|$ of around $2\%$, although we investigate larger $|\kappa|$ in Section \ref{realistic}. The $\kappa$-field is generated on the 100x100 grid, each point representing a postage stamp, and is applied to the galaxy at the same position ($s=s^s(1+\kappa)$), neglecting the contribution of the shear. The pixel angular size is not fixed, and can be scaled to any experimental set up.  

\begin{table*}
\centering
\begin{tabular}{l c c c c c c c}
Set Name & S/N & \textit{fwhm} PSF(pix) & \textbf{e} PSF & B/D fraction & $r_{d}$(pix) & $r_{b}$(pix) &{\bf e}\\\hline \hline 
Set 1 & \textbf{[10,40]} & 0.01 & 0 & 0 & $\langle r_d\rangle=7$ , $\sigma=1.2$ & - & 0 \\\hline
Set 2 & 20 & \textbf{[0.01,10]}& 0 & 0 & $\langle r_d\rangle=7$ , $\sigma=1.2$ & -& $\langle{\bf e}\rangle=0$,$\sigma=0.3$ \\\hline
Set 3 & 20 & 0.01 & 0 & \textbf{[0,0.95]} & $\langle r_d\rangle=7$ , $\sigma=1.2$ &$r_{d}/2$  &$ \langle{\bf e}\rangle=0$,$\sigma=0.3$ \\\hline
Set 4 & \textbf{[10,40]} & 4.5 & [-0.1,0.1] & 0.5 & $\langle r_d\rangle=7$ , $\sigma=1.2$ & $\langle r_b\rangle=3.5$ , $\sigma=0.6$ & $\langle{\bf e}\rangle=0$,$\sigma=0.3$ \\\hline
\end{tabular}
\caption{\small{ Major characteristics of the different sets used in this analysis. In bold are marked the variables explored in each set and the range of variation.	In Set 4, PSF ellipticities are drawn from a uniform distribution in the range specified. Note that $r_i$ corresponds to the half-light radius. Last column is the galaxy ellipticity and is the same for both components, bulge and disk.}}
\label{table1}
\end{table*}

\section{Results}
Before trying to estimate the convergence field of our most realistic image, we have tested the dependence on different aspects separately: 
S/N, PSF size and bulge fraction. We expect these to be the observable effects that have the largest impact on the ability to measure the 
size of galaxies. Lower S/N will cause size estimates to become more
noisy and possibly biased (in a similar way as for ellipticity, see
\citealt{Melchior2012}); a larger PSF size will act to remove information on galaxy size from the image and a change in galaxy type or bulge fraction may cause biases because 
now two characteristic sizes are present in the images (bulge and disk lengths). 
In order to study carefully the sensitivity of our estimator to systematic noise, PSF or galaxy properties, we started from the simplest 
case and added increasing levels of complexity. The number of galaxies used for the analysis is 200,000 for the first three sets and 
we increased the number to 500,000 for the last test to give smaller error bars.

\begin{figure*}
\centering
\resizebox{13cm}{!}{\includegraphics{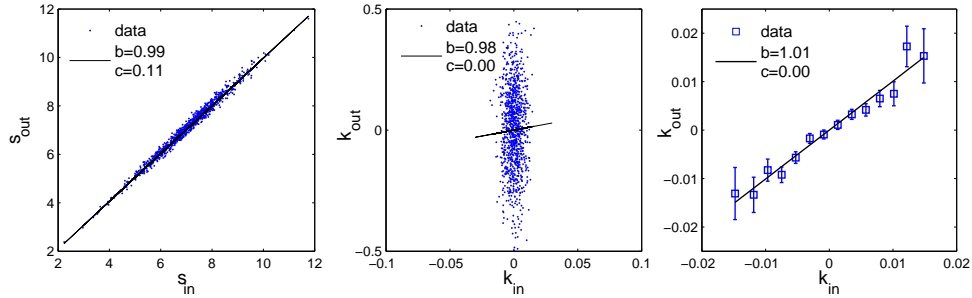}} 
\caption{Sequence of steps to obtain $m$ and $c$ values. First panel shows the {\em lens}fit output size compared to the input size, in the second panel the estimated $\kappa$ compared to the input convergence at each galaxy, and in the third panel is shown the same plot using bins. Slope and intercept values of the fitting are shown in each plot (throughout, we fit generically $y=bx+c$, with $b=m+1$ and $c=c$ of eq.~\ref{mc}). This is for galaxies of Set 1 with signal-to-noise 40.}
\label{three_panel}
\end{figure*}

We compare the estimated $\hat{\kappa}$ computed as in eq.~\ref{estimator} with the input field $\kappa$, and fit a straight line to the relationship to estimate a multiplicative bias $m$ and an additive bias $c$:  
\begin{equation}
\hat{\kappa}=(m+1)\kappa+c,
\label{mc}
\end{equation}
after applying the sensitivity correction (section~\ref{sens}).   We bin the data\footnote{Note that the differences between the results before and after the binning of $\kappa$ are within the error bars.} and a linear fit is done to compute $m$ and $c$. This process is shown in Fig.~\ref{three_panel}. The error bars for the regression coefficients are given by its standard deviation, assuming that the errors are normally distributed. The regression coefficients and its errors are computed using the function {\it polyfit} of {\sc matlab} software. 

We now discuss each of the categories in turn. 

\subsection{Signal-to-noise}
\begin{figure}
\centering
\resizebox{7.5cm}{!}{\includegraphics{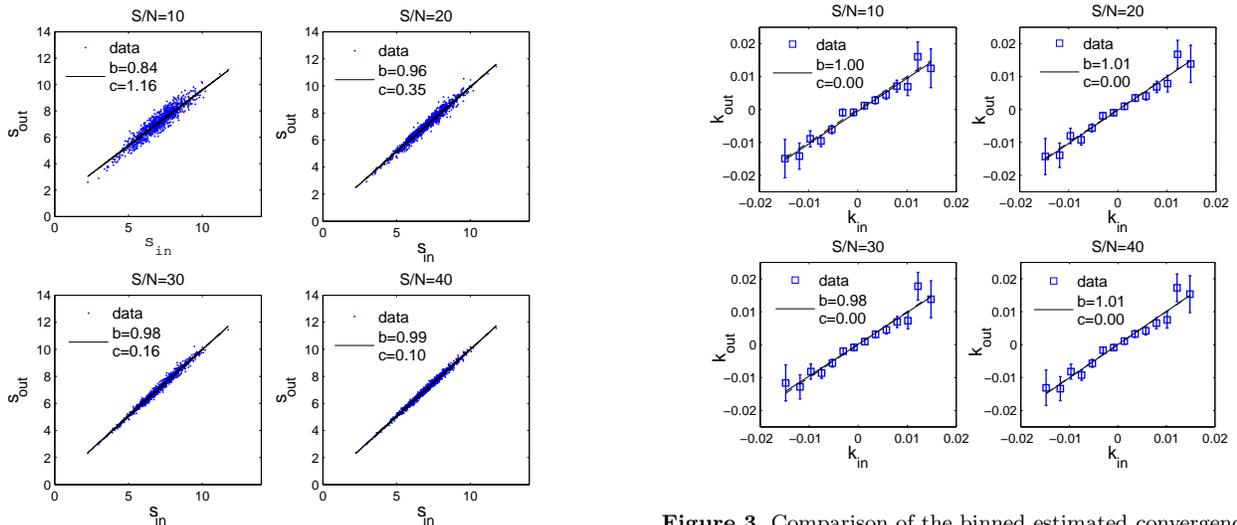}} 
\caption{Comparison of the estimated sizes by {\em lens}fit with the input galaxy size for different S/N in the range [10,40]. Disk-only circular galaxies with a negligible PSF effect are considered (Set 1). Slope and intercept of the fitting are shown (b and c, respectively). Note that the input size is the lensed one.}
\label{sizes_bf0_SN}
\end{figure}

\begin{figure}
\centering
\resizebox{7.2cm}{!}{\includegraphics{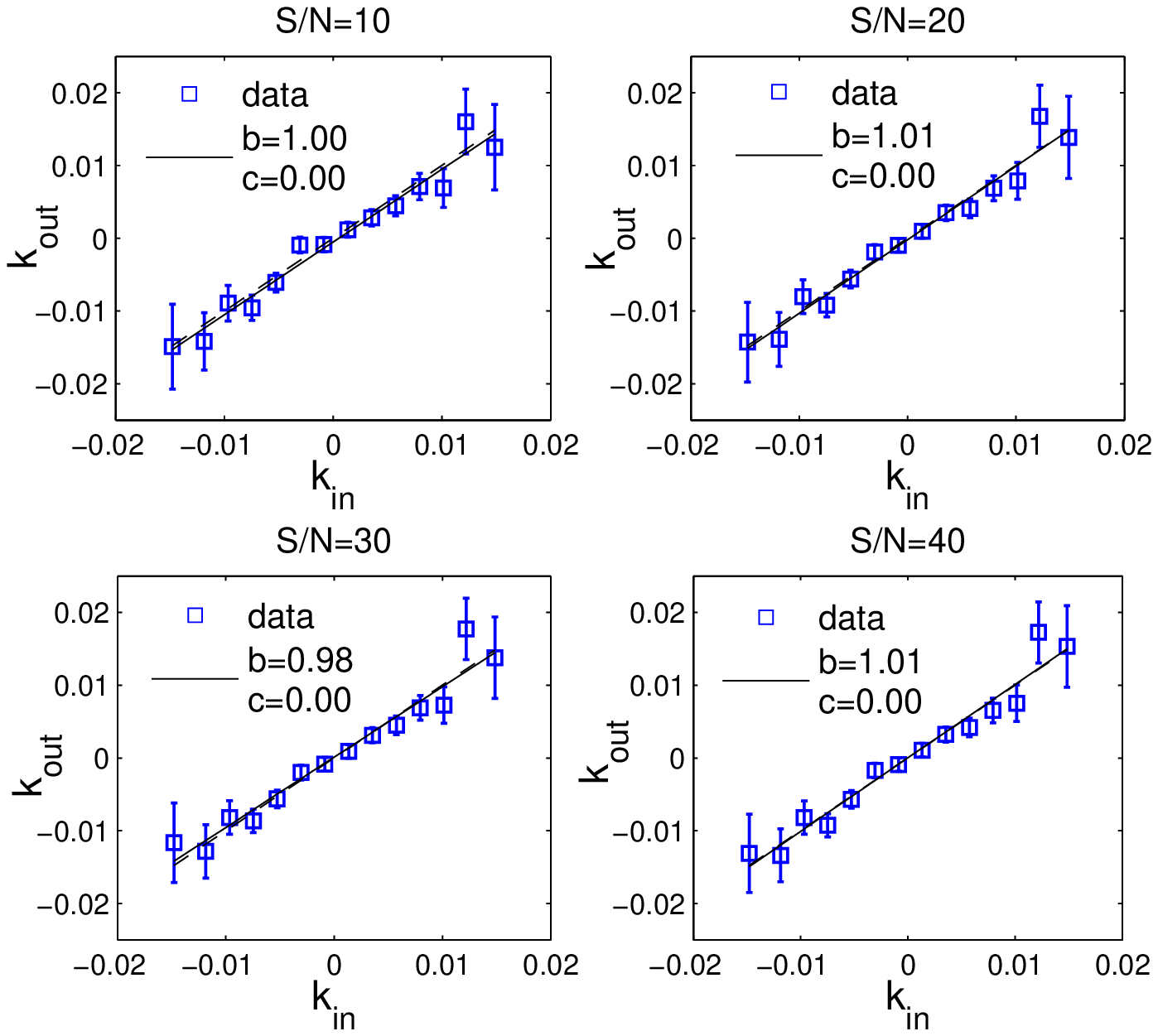}} 
\caption{Comparison of the binned estimated convergence and the input value for Set 1 with different S/N in the range [10,40]. Slope and intercept of the fitting are shown (b and c, respectively). For errors, see text.}
\label{kappa_bf0_SN2}
\end{figure}

\begin{figure}
\centering
\resizebox{8.5cm}{!}{\includegraphics{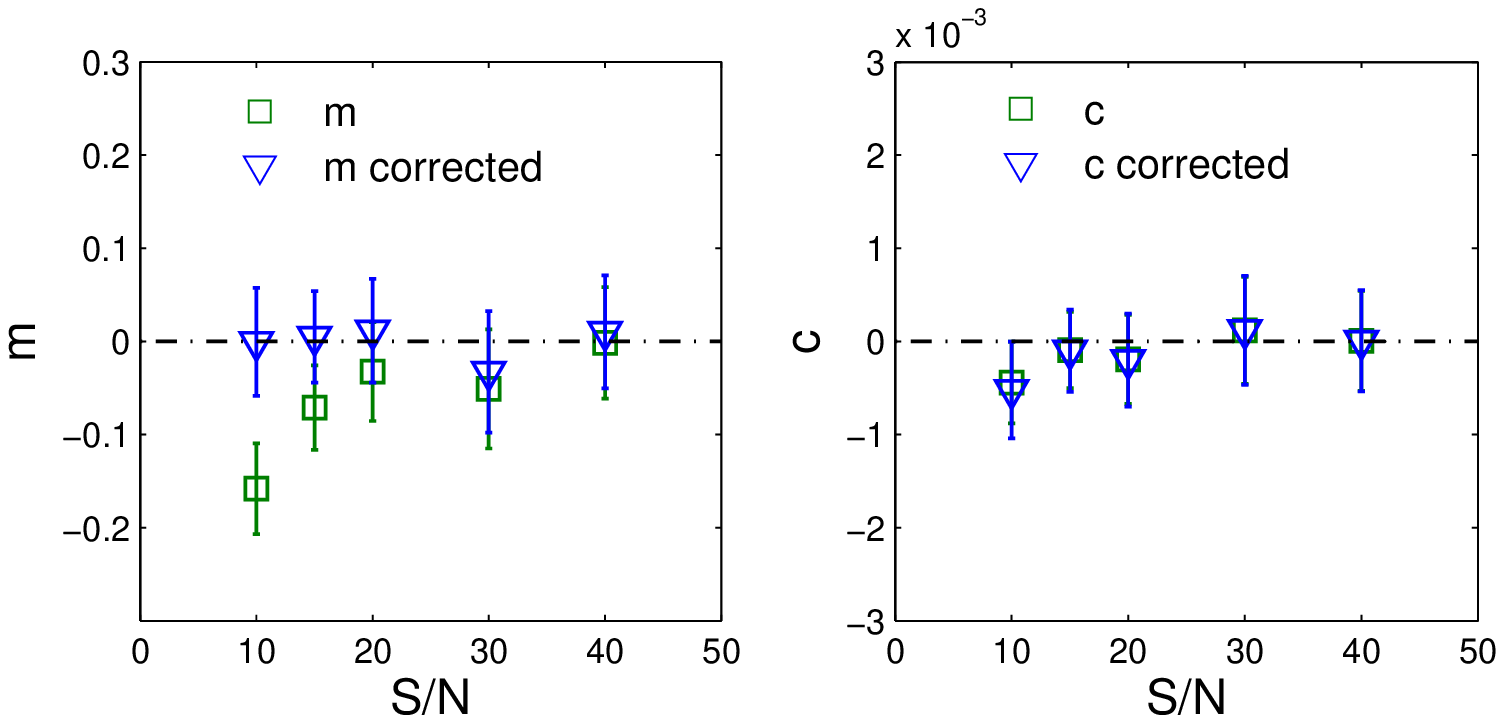}} 
\caption{$m$ and $c$ values computed with 200,000 galaxies of Set 1. Triangles are for the values obtained with the sensitivity correction and squares without it.}
\label{m_c_bf0_SN}
\end{figure}

As a first approach to the problem, disk-only galaxies with a negligible PSF (FHWM$=0.01$ pixels) and zero ellipticity  
were generated to test the 
dependence of the bias on S/N alone, given otherwise perfect data. In Fig.~\ref{sizes_bf0_SN} we can see that as we increase the S/N the accuracy of the size estimation grows, as expected. 
In Fig.~\ref{kappa_bf0_SN2} we show the estimation of the convergence field, modified by the sensitivity correction. There is a clear correlation between the inputs and the outputs, and the slope is close to unity for all S/N explored.
In Fig.~\ref{m_c_bf0_SN} the estimates for $m$ and $c$ are shown, with and without the sensitivity correction. In this case the correction 
does not alter the results much except at low S/N, because the sizes are less accurately estimated. 
In this paper the factor $a^2/(a^2+b^2)$ is estimated by the inverse of the slope of the size estimation fitting (see Fig.~\ref{sizes_bf0_SN}). 
Using 200,000 galaxies for this test, the values found for $m$ and $c$ are consistent with zero, typically $m\simeq0.02\pm 0.05$, and $c \simeq (5\pm 5) \times10^{-4}$. 

\subsection{PSF effect}
\begin{figure}
\centering
\resizebox{7.5cm}{!}{\includegraphics{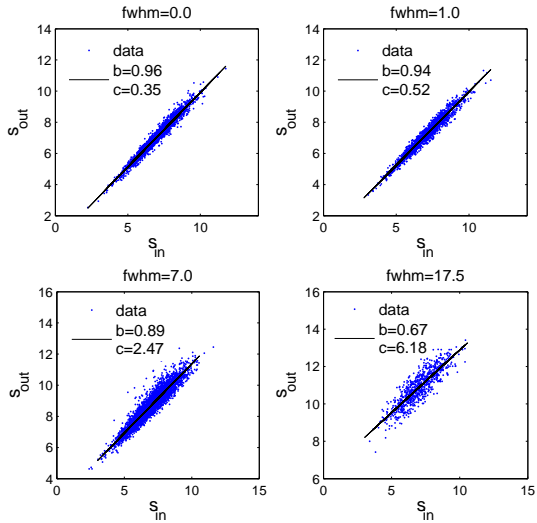}} 
\caption{Sizes estimates vs input sizes for four different PSF scale-lengths between 0.1 and 7 pixels. Galaxies are disks with S/N=20 and mean size 7 pixels. Slope and intercept of the fitting are shown (b and c, respectively).}
\label{sizes_bf0_PSF}
\end{figure}

\begin{figure}
\centering
\resizebox{7.7cm}{!}{\includegraphics{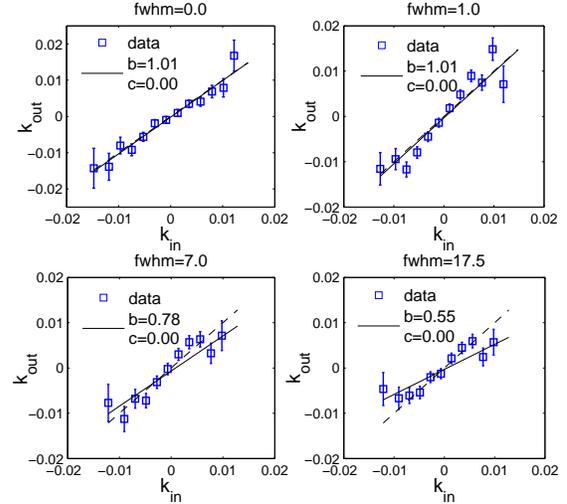}} 
\caption{$\kappa$ estimates vs input values for four different PSF scale-lengths. Galaxies are disks with S/N=20 and mean size 7 pixels. Dashed line is $\kappa_{out}=\kappa_{in}$ and the solid line is the least squares fit, with slope and intercept shown in the plots. Note that b=m+1.}
\label{kappa_bf0_PSF}
\end{figure}

\begin{figure}
\centering
\resizebox{8.5cm}{!}{\includegraphics{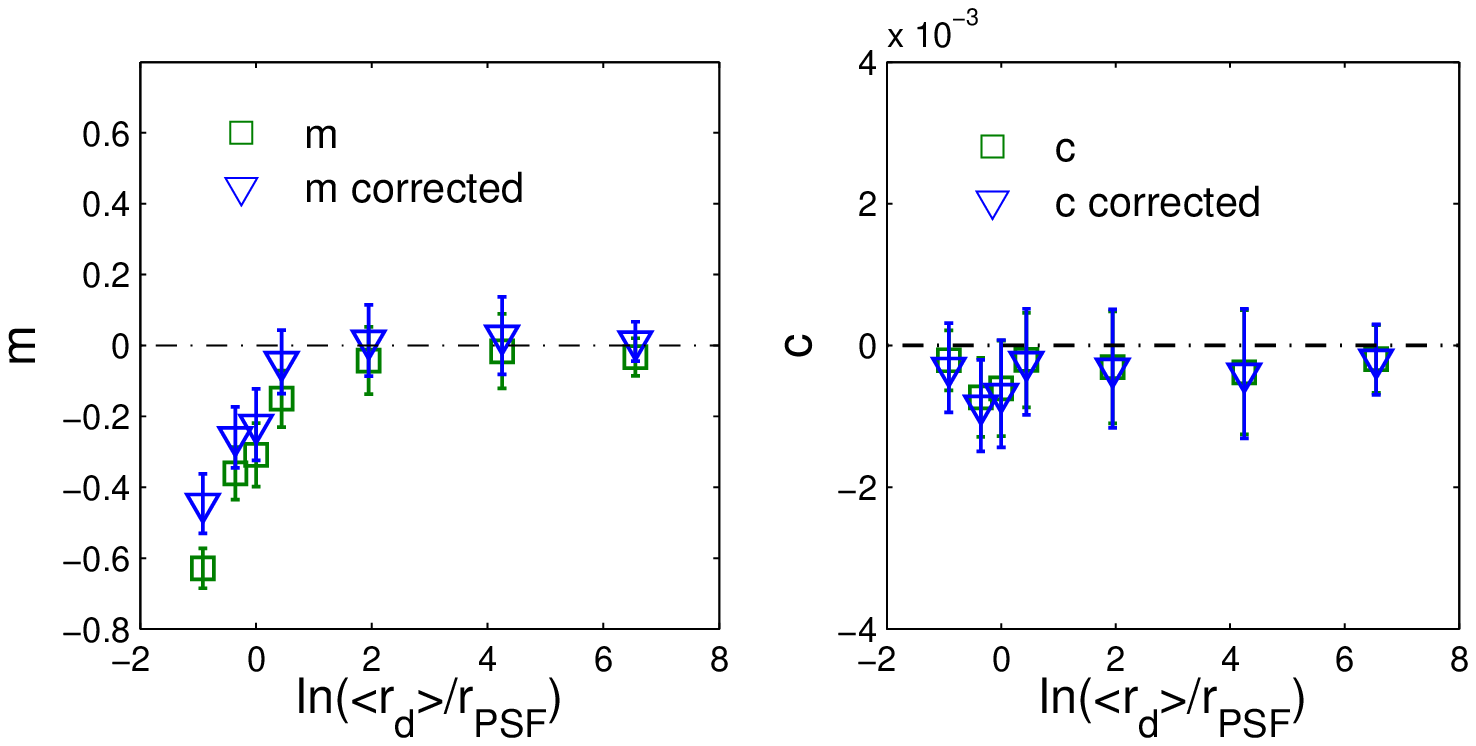}} 
\caption{$m$ and $c$ values computed with 200,000 galaxies of Set 2. Triangles represent the values obtained with the sensitivity correction and squares without it.}
\label{m_c_bf0_PSF}
\end{figure}

To study the uncertainties on the size estimation due to the PSF size, we generated images with different FWHM PSF values, with an intermediate signal to noise (S/N=20), maintaining the same properties as before, except that we considered here a Gaussian distribution of ellipticities with mean value of $e=0$ and $\sigma_{e}=0.3$ (per component). 
The size estimates are good for small PSFs, but become progressively more biased as the PSF size increases beyond the disk scale length (see Fig.~\ref{sizes_bf0_PSF}). A PSF with a FWHM larger or similar 
to the size of the disk, tends to make the galaxy look larger, and the estimator for $\kappa$ becomes biased. This effect can be seen in the slope and intercept of $\hat{\kappa}$ vs $\kappa$ 
plot (Fig.~\ref{kappa_bf0_PSF}).
Fig.~\ref{m_c_bf0_PSF} shows the variation of the parameters $m$ and $c$ with the ratio between the scale-length of the PSF and the galaxy 
(ratio=$r_{d}/PSF_{\rm FWHM}$). 

We find no evidence for an additive bias, but we do find a multiplicative bias for large PSFs.  
With a wide size distribution, some of the smaller galaxies are convolved with a PSF larger than their size, and this could produce an  
overall bias in $\kappa$, but if the number of those galaxies is not very large, the effect on the global estimation of the convergence field will be correspondingly small. 
Similar biases exist with shear measurement for large PSFs, but the biases are
larger here. For a space-based experiment, with a relatively bright
cut at $i\sim 24.5$, such as planned for Euclid, the limitation on PSF
size will not be dominant because the median galaxy size is 0.24 arcsec
\citep[]{Simard2002,Miller2012}, larger than the PSF FWHM of 0.18 arcsec.
For ground-based surveys, such as CFHTLenS and future experiments the
situation is not so clear, the measurement will be more challenging, and large empirical bias corrections of the order of $m\simeq -0.5$ will be needed
(see the first point of Fig.~\ref{m_c_bf0_PSF}).

\subsection{Bulge fraction}
\begin{figure}
\centering
\resizebox{8.3cm}{!}{\includegraphics{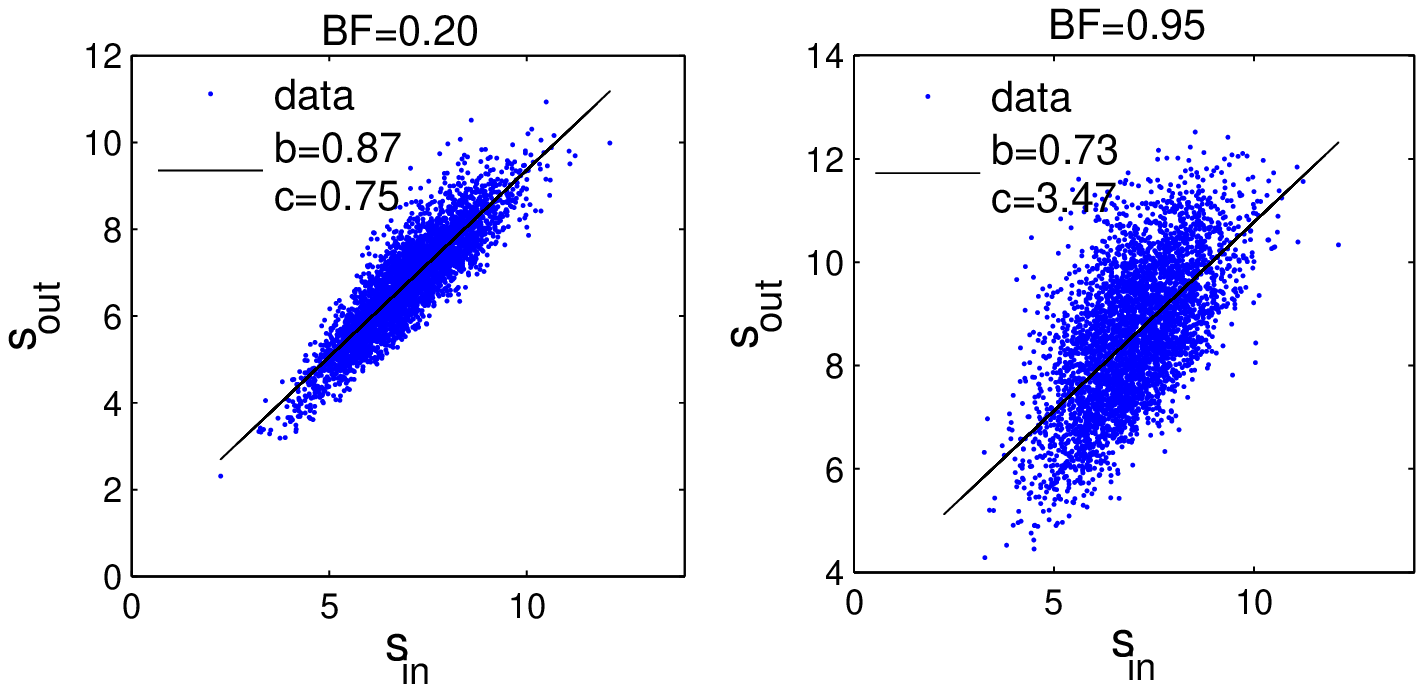}} 
\resizebox{8.5cm}{!}{\includegraphics{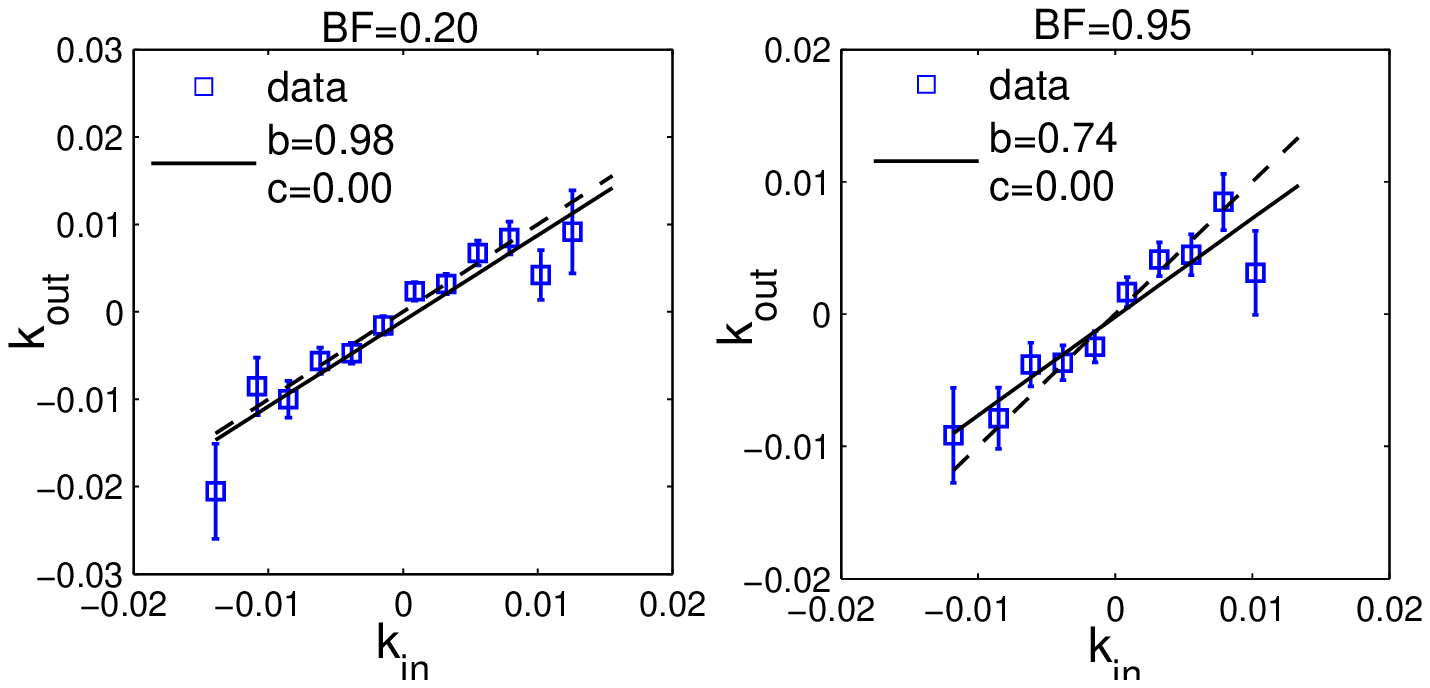}} 
\caption{Top panels: Sizes estimates vs input sizes for different bulge fractions in the range [0.2,0.95]. Bulge+Disk galaxies with different ellipticities are used, with S/N=20 and negligible effect of the PSF (Set3). Slope and intercept of the fitting are shown (b and c, respectively). Bottom panels: Bulge+Disk elliptical galaxies with S/N of 20 and negligible effect of the PSF (Set 3). Dashed line is $\kappa_{out}=\kappa_{in}$ and the solid line is the fit of the output values. Note that b=m+1 of eq.~\ref{mc}.}

\label{sizes_bf_PSF001}
\end{figure}
\begin{figure}
\centering
\resizebox{8.5cm}{!}{\includegraphics{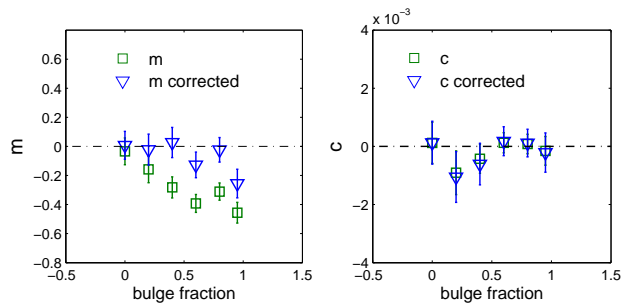}} 
\caption{$m$ and $c$ values computed with 200,000 galaxies of Set 3. Values obtained with the sensitivity correction are marked by triangles and by squares are without the correction.} 
\label{m_c_bf_PSF001}
\end{figure}

In this test we generated galaxy images with bulges with different fractions of the total flux, to test the response to the galaxy type.  In Fig.~\ref{sizes_bf_PSF001} we show that galaxy size estimates for bulge fractions of 0.2 are much better than for galaxies with bulge fractions of 0.95.  
This is because for bulge-dominated models the central part of the galaxy becomes under-sampled due to a limiting pixel scale. The poor estimation of sizes is reflected in the $\kappa$ estimation (bottom panels of Fig.~\ref{sizes_bf_PSF001}).
The parameters $m$ and $c$ for this set are shown in Fig.~\ref{m_c_bf_PSF001}, where we can see that for bulge fraction greater than 0.8, the results are clearly biased with $m=-0.25$. For all bulge fractions the error bars are around 10\%. Although for bulge-only galaxies, the $\kappa$ estimates are poor most of the galaxies used for weak lensing experiments have bulge fractions lower than 0.5 \citep{Schade1996}, so that in fact the population of useful lensing galaxies is likely to be enough to do a successful analysis.

\FloatBarrier

\subsection{Most Realistic Set}
\label{realistic}
\begin{figure*}
\centering
\mbox{\subfigure{\includegraphics[width=3in]{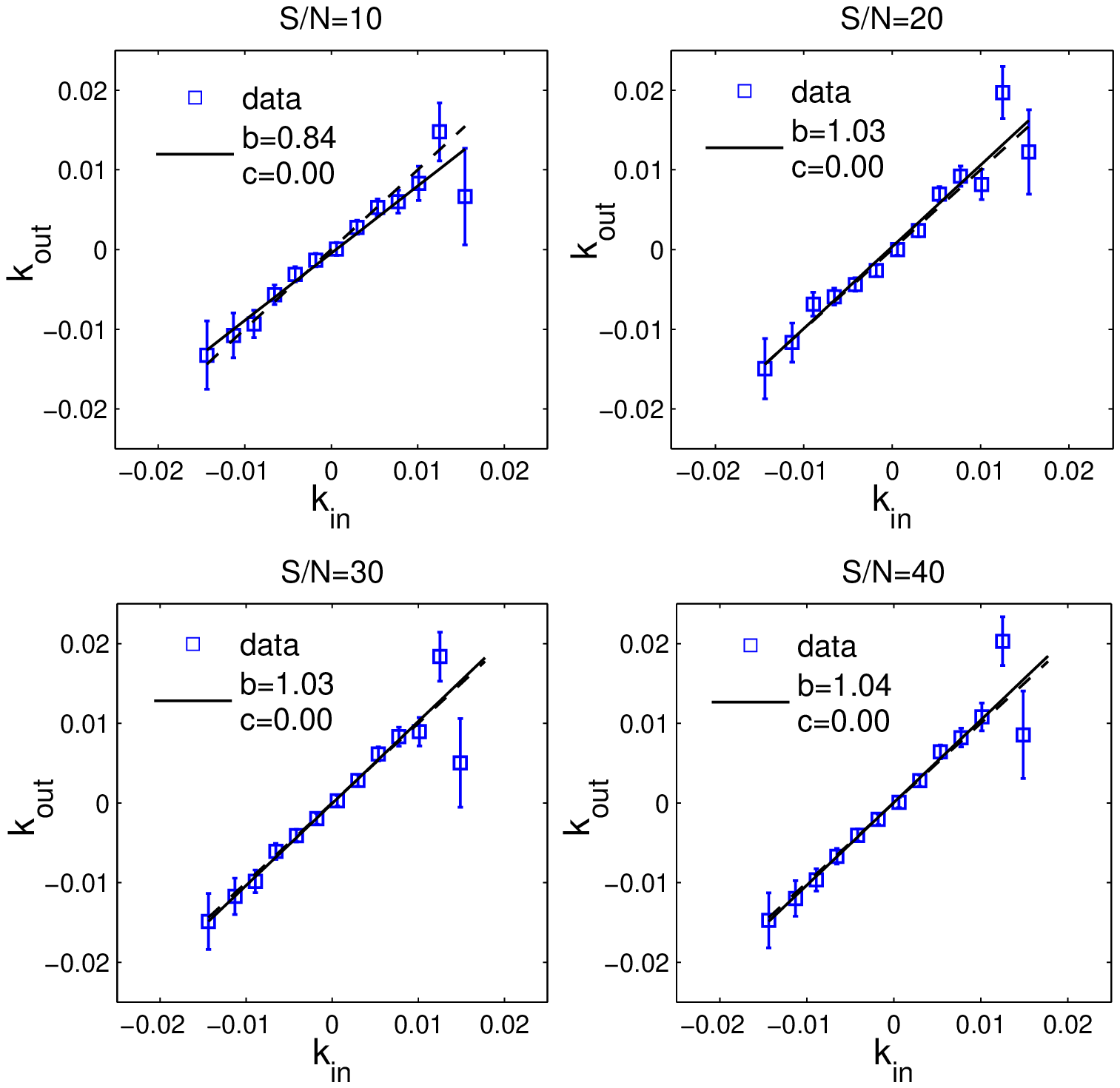}}
\quad
\subfigure{\includegraphics[width=3in]{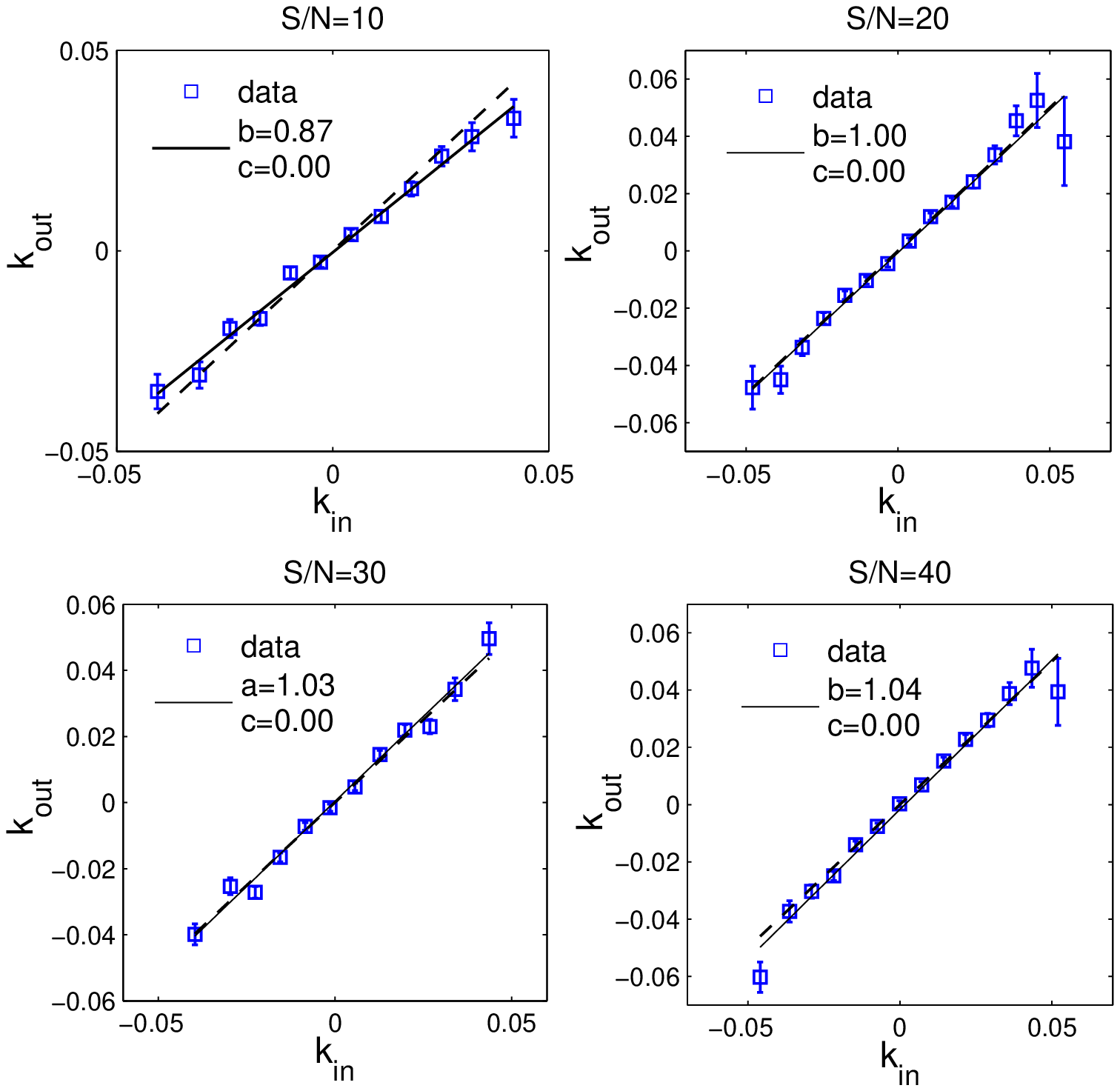} }}
\caption{$\hat{\kappa}$ estimates compared to the true $\kappa$ values for set 4. Dashed line is for $\kappa_{in}=\kappa_{out}$ and solid line is the least squares fit, with regression coefficients shown in the plot ($\hat{\kappa}=b\kappa+c$, where $b=m+1$). Left four plots are for a maximum value of input $\kappa$ of $\sim 2\%$ and right set of plots for $|\kappa_{in}|\lesssim0.05$.} \label{comparison}
\end{figure*}

\begin{figure}
\centering
\resizebox{8.5cm}{!}{\includegraphics{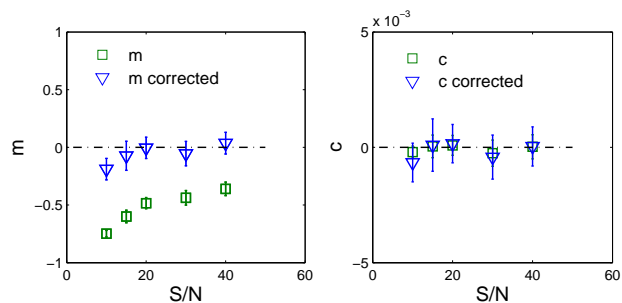}} 
\caption{$m$ and $c$ parameters for 500,000 galaxies of Set 4. Squares are raw $m$ and $c$  values; triangles have the sensitivity correction included. }
\label{m_c}
\end{figure}

The last set includes realistic values for all the effects we investigate. We have generated galaxies with elliptical isophotes with a bulge fraction of 0.5 convolved with 
an anisotropic PSF with FWHM of 4.5 pixels (1.5 times smaller than the characteristic scale-length of the disk), and again we investigate the dependence on S/N.  This is also a challenging test for {\em lens}fit, the current version (c. 2012) of which uses a simplified parameter set where the bulge scale length is assumed to be half the disk scale length. Here, we include a dispersion in the bulge scale length of 0.6 pixels around a mean value of 3.5 pixels. The analysis was done with 500,000 galaxies, to keep the error bars smaller than 10\%. As expected, the size estimation for this set is poorer than in set 1; however, the errors on $\hat{\kappa}$ remain similar thanks to the sensitivity correction (see Fig.~\ref{comparison}).
While the theoretical $\kappa$ in the studied range of $\ell$ has a maximum amplitude of $\sim2\%$,  non-linear density evolution increases its contribution on smaller scales \citep[see for example Fig.17 of][]{Bartelmann1996}. For this set, the analysis was performed with higher values of $\kappa$ to confirm that the method is valid for larger $\kappa$. A comparison of the original range with a larger range ($|\kappa|\lesssim0.05$) is shown in Fig.~\ref{comparison}, where no significant differences are apparent. 
Fig.~\ref{m_c} shows the values of $m$ and $c$ for this set, with and without the sensitivity correction. If we compare it with 
the previous plots we can see that as the galaxy population becomes more realistic, including several effects, the importance of the correction increases. 
Results for this set are shown in Fig.~\ref{m_c}, showing unbiased results except for S/N=10, which has $m=-0.19 \pm 0.1$. For the higher S/N points, we find $|m|<0.06$ with errorbars of $\pm 0.09$.

\section{Conclusions and discussions}
In this paper we present the first systematic investigation of the performance of a weak lensing shape measurement method's ability to estimate the 
magnification effect through an estimate of observed galaxy sizes. We performed this test by creating a suite of simulations, with known input values, 
and by using the most advanced shape measurement available at the current time, {\em lens}fit. 
 
A full study of the magnification effect using sizes was performed testing the dependence on S/N,  PSF size and type of galaxy. 
The requirements on biases on shear (or equivalently convergence) for
Euclid such that systematics do not dominate the very small statistical
errors in cosmological parameters are stringent (see Massey et al., 2012 and Cropper et al., 2012). A much larger study
will be required to determine whether these requirements can be met
for size, but we find in this study no evidence for additive size biases at all, and no evidence for multiplicative bias provided that 1) 
the PSF is small enough ($< $galaxy scale-length/1.5), 2) the S/N high enough ($\geq 15$), and 3) the bulge not too dominant (bulge/disk ratio $<=4$).

Besides instrumental and environmental issues, there can be astrophysical contaminants 
associated with weak lensing. In the case of shape distortion, the first assumption that galaxy pairs have no ellipticity correlation is not entirely accurate. 
There are intrinsic alignments of nearby galaxies due to the alignment of angular momentum produced by tidal shear 
correlations (II correlations, see for detections \citet{Brown2002,Heymans2004,Mandelbaum2011,Joachimi2011,Joachimi2012}, and for theory \citet{Heavens2000,Catelan2001,Crittenden2001,Heymans2003}). 
In addition there can be correlations between density fields and ellipticities 
\citep[GI correlations][]{Hirata2004,Mandelbaum2006}. These
intrinsic correlations have been studied in detail and it is not trivial to account for or remove 
them when quantifying the weak lensing signal. 
\FloatBarrier

The intrinsic correlation of sizes and its dependence on the environment, are still open issues. In fact, the correlation of sizes and density field, 
is known to play an important role in discriminating between models of size evolution; recent work finds a significant correlation between sizes 
and the density field using around $11$,$000$ galaxies drawn from the joint DEEP2/DEEP3 data-set \citep{Cooper2012, Papovich2012}, 
while earlier studies with smaller samples have been in disagreement. Using $5$,$000$ galaxies of STAGES data-set, \citet{Maltby2010} find 
a possible anti-correlation between the density field and size for intermediate/low-mass spiral galaxies. Clustered galaxies seem to be $15\%$ 
smaller than the ones in the field, while they do not find any correlation for high-mass galaxies. Also for massive elliptical 
galaxies from ESO Distant Clusters Survey, \citet{Rettura2010} do not find any significant correlation, while using the same 
data set \citet{Cimatti2012} claims a similar correlation as in \citet{Cooper2012}. 
In \citet{Park2009} they study the correlation between sizes and separation with late and early-type galaxies from SDSS catalogue, 
at small and large scales. They compare the size of the nearest neighbour with the separation between them, and find larger galaxies at 
smaller separations. This correlation is found for early-type galaxies if the separation between the galaxies is smaller than the merging 
scale, but not for larger separations. The size of late-type galaxies seems not to have a correlation with the separation in any scale.
We expect that further studies with larger samples will clarify the intrinsic correlations of sizes, we note that the systematics 
are generated from different physical processes than in the case of
shear and so will affect the signal in a different way; we suggest
this is a positive, and another reason why a joint
analysis of ellipticity and sizes is interesting. 

The analysis presented here has assumed that the statistical distribution of galaxy sizes is known, whereas in practice the size distribution depends on galaxy brightness and must be determined from observation.  Gravitational lensing of a galaxy with amplification $A$ increases both the integrated flux and area of that galaxy by $A$, which has the effect
of moving galaxies along a locus of slope 0.5 in the relation between log(size) and log(flux). Thus, if the intrinsic distribution of sizes $r$ of galaxies scales with flux $S$ as $r \propto S^\beta$, the apparent shift in size caused by lensing amplification $A$ is $r' \propto A^{0.5-\beta}$, resulting in a dilution of the signal compared with the idealised case investigated in this paper.  A similar effect occurs in galaxy number magnification, where the observed enhancement in galaxy number density $N'$ varies as $N' \propto A^{\alpha-1}$, if the intrinsic number density of galaxies varies as $N \propto S^{-\alpha}$ \citep{Broadhurst1995}. The value of $\beta$ at faint magnitudes has recently been estimated by \citet{Miller2012}, who analysed the fits to galaxies with $i \la 25$ of \citet{Simard2002} and estimated $\beta \simeq 0.29$.  Thus we expect this effect in a real survey to dilute the lensing magnification signal by a factor 0.42, but still allowing detection of lensing magnification.  In practice, the dilution factor could be evaluated by fitting to the size-flux relation in the lensing survey.

Lensing number magnification surveys are also affected by the problem that varying Galactic or extragalactic extinction reduces the flux of galaxies and thus may cause a spurious signal \citep[e.g.][]{Menard2010}.  Such extinction would also affect the size magnification of galaxies, but with a different sign in its effect.  Thus a combination of lensing number magnification and size magnification might be very effective at removing the effects of extinction from magnification analyses.

Space-based surveys as Euclid should overcome the limitations that we have exposed here, having a large number of galaxies, with S/N $>$ 10, and importantly
a PSF at least 1.5 smaller than the average disk size.
The addition of the size information to the ellipticity analysis is
expected to reduce the uncertainties in the estimation of
weak lensing signal, and therefore improve the constraints of the distribution of matter and dark energy properties.

\section*{acknowledgments}
We thank Catherine Heymans for interesting discussions. 
BC thanks Chris Duncan for useful comments on the first draft. BC thanks the Spanish Ministerio de Ciencia e Innovaci\'on for a pre-doctoral
fellowship. We acknowledge partial financial support from the Spanish Minsterio De Econom{\'\i}a y Competitividad AYA2010-21766-C03-01
and Consolider-Ingenio 2010 CSD2010-00064 projects.TDK is supported by a Royal Society University Research Fellowship. The authors acknowledge the computer resources at the Royal Observatory Edinburgh. 
The author thankfully acknowledges the computer resources, technical expertise and assistance provided by the Advanced Computing \& e-Science team at IFCA.
\bibliographystyle{mn2e}
\bibliography{weak_lensing_mag_rev2_v2_astroph}

\end{document}